%% file: main.tex
\documentclass[12pt, english, a4paper]{scrarticle}

\usepackage{amssymb}
\usepackage{libertine}  
\usepackage{libertinust1math}  
\usepackage[scale=0.92, varqu]{inconsolata}  
\usepackage{microtype}

\usepackage[a4paper, left=3.5cm, right=3cm, top=3cm, bottom=3cm]{geometry}

\usepackage{setspace}
\usepackage{authblk}

\usepackage{graphicx}
\graphicspath{assets/}

\usepackage{array}
\usepackage[frozencache, cachedir=.]{minted}
\usepackage{booktabs}
\usepackage{multirow}
\usepackage{siunitx}

\usepackage{xcolor}
\definecolor{MaRDIBlue}{HTML}{005eaa}      
\definecolor{MaRDIOrange}{HTML}{d0662b}     

\usepackage[
  colorlinks = true,
  linkcolor=MaRDIBlue,
  citecolor=MaRDIBlue,
  urlcolor=MaRDIOrange,
]{hyperref}

\input{macros}

\usepackage[
  natbib=false, style=numeric, giveninits=true, uniquename=init,
  isbn=false
]{biblatex}
\title{MaRDI Open Interfaces\\for Interoperable\\Nonlinear Optimization}
\author{Dmitry I.\ Kabanov, Stephan Rave, Mario Ohlberger} 
\date{\small 18 June 2026}
\affil{\small \emph{Mathematics Münster, University of Münster, Germany}}
\newcommand{\myabstract}{%
  MaRDI Open Interfaces is a software package that aims to improve
  interoperability in scientific computing, particularly,
  for nonlinear optimization.
  To this end, this package holds two main characteristics.
  First, it provides unified interfaces
  for typical numerical problems
  to help switching between solvers for the same problem type.
  Second, it automates data marshalling between programming languages.
  Hence, computational scientists can conduct experiments faster
  by using the package, with fewer code-modification and testing efforts.
  In this work we describe the general structure of the software package
  and show examples with the interface for nonlinear optimization.%
}
\newcommand{\keywords}{unified solver interfaces; numerical software; scientific computing; computational science; cross-language computations; automatic data exchange; data marshalling; programming interface; reusability; foreign function interface}

\newenvironment{acknowledge}
{\paragraph{Acknowledgements}}
{}

\begin{document}
\maketitle
\begin{abstract}
    \myabstract{}
\end{abstract}

\input{01-introduction}
\input{02-description}
\input{03-included-interfaces}
\input{04-examples}

\begin{acknowledge}
This work was funded by Nationale Forschungsdaten Infrastruktur
(NFDI, National Research Data Infrastructure),
project number~460135501, NFDI~29/1
“MaRDI – Mathematical Research Data Initiative
[Mathematische For\-schu\-ngs\-da\-teninitiative]”~\cite{BennerEtAl2022}
and additionally by the Deutsche Forschungsgemeinschaft (DFG, German Research Foundation) under Germany's Excellence Strategy EXC 2044/2 - 390685587, Mathematics Münster: Dynamics-Geometry-Structure.
\end{acknowledge}

\printbibliography{}

\end{document}

%% file: macros.tex
\usepackage{amsmath}

\newcommand{\R}{\mathbb{R}}

\DeclareMathOperator*{\argmin}{arg\,min}

\newcommand{\norm}[1]{\lVert #1 \rVert}


%% file: 01-introduction.tex
\section{The need for increased interoperability in Scientific Computing}

Building complex computational experiments that consists of multiple
software packages for scientific computing,
often encounters obstacles
arising from the heterogeneity of these packages.
Particularly, when a researcher writes experiment's code
in one programming language
and needs to use a solver written in another,
this researcher must write bindings to this solver
using specialized tools
or rely on existing bindings if they are provided by the package authors.
Another obstacle occurs when benchmarking different packages
(in terms of runtime performance or obtained solutions)
aimed at solving the same computational problem,
as this necessitates adapting to different programming interfaces
used by the packages.
Both of these obstacles significantly increase coding and testing efforts
required to conduct computational experiments.
Furthermore, such efforts are often repeated across different projects,
as they often stay bound to a particular project.

\emph{MaRDI Open Interfaces} is a software package designed
to address these obstacles and increase interoperability in scientific computing.
To achieve this goal, the package provides unified interfaces
for various numerical solvers that solve the same problem type,
such as the numerical integration of differential equations or nonlinear optimization.
Additionally, the library provides automated data marshalling
between different programming languages.
In that sense, the aim of the package is similar
to other projects such as \emph{preCICE}~\cite{Chourdakis2022}
for coupling finite-element and finite-volume solvers in multi-physics simulations
and \emph{UM-Bridge}~\cite{SeelingerEtAl2023}
for uncertainty quantification (UQ) problems where models and UQ algorithms
are loosely coupled and can be written in different languages.

This software package is part of the
\emph{Mathematical Reseach Data Initiative} (MaRDI)~\cite{BennerEtAl2022},
a nationwide German project for mathematical sciences
by the \emph{National Research Data Infrastructure} (NFDI).

%% file: 02-description.tex
\section{Description of \emph{MaRDI Open Interfaces}}

In this section we give a short overview of the implementation details
of \emph{MaRDI Open Interfaces}.
More details on the package can be found in~\cite{KabanovEtAl2025}.

\subsection{Software Architecture}
The architecture of the package is based on decoupling of the user side
and the implementation side by organizing them as a set
of dynamically loaded libraries
that interact via predefined interfaces,
following well-known principle ``\emph{program to an interface,
  not an implementation}''~\cite[p. 18]{GammaEtAl1995},
which is used multiple times:
first, to decouple user's code from the details of the package itself,
second, to decouple internal dispatching library from the details
of runtimes of the different programming languages,
and third, to conform existing implementations to the same interfaces
that are used by the user.
Additionally, the data passing uses conversion of the data types
in an intermediate C representation, as, in general,
programming languages used in scientific computing
have functionality to communicate with C data types and libraries.
Third, all communication and data passing happens in-process,
that allows avoiding performance penalties, as data are passed
by pointers, which is especially important for arrays
as their memory size is defined by the problem at hand,
and copying such data structures would be costly.

\begin{figure}
  \centering
  \includegraphics[width=\textwidth]{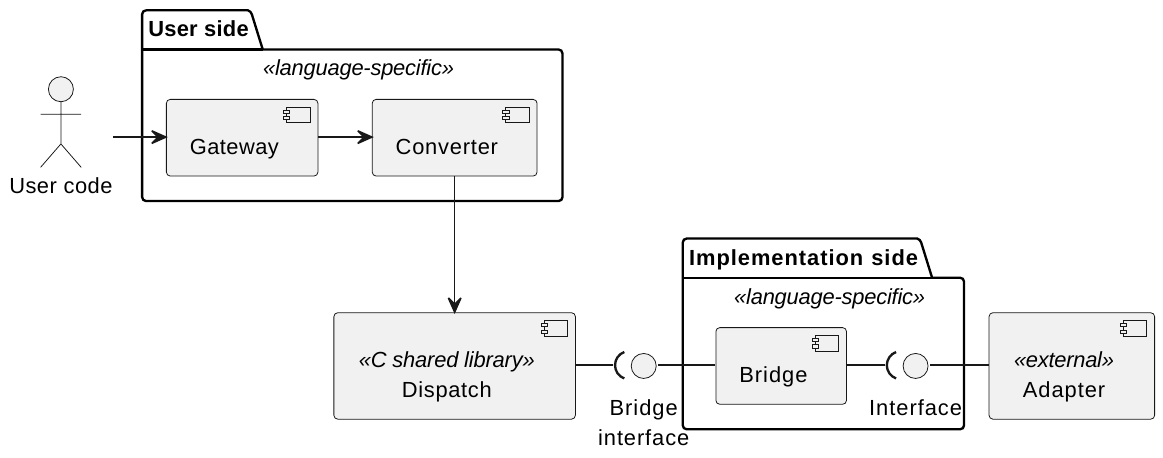}
  \caption{
    Component diagram of \emph{MaRDI Open Interfaces}.\label{fig:components}}
\end{figure}

The software package consists of components of different types,
which, together with external components,
are shown on the component diagram in~\autoref{fig:components}.
The system overall is split in packages:
the user side and the implementation side, each of which
can be implemented in different languages.
They communicate with each other via a shared library \emph{Dispatch}.
The roles of the components are the following:
\begin{enumerate}
  \item \emph{Gateway} provides a unified interface
        to different solvers for a particular numerical problems.
        Such a component exists for each supported language and for each
        interface.
        To instantiate a gateway, the user needs to specify the required
        implementation as a string identifier.
  \item \emph{Converter} converts data to intermediate C data types.
        Such a component exists for each supported language.
  \item \emph{Dispatch} provides the following operations.
        It finds implementation details of the requested implementation
        from configuration files on disk,
        loads the necessary \emph{Bridge} (described below) and passes
        the implementation details to the Bridge.
        After initialization of an implementation, \emph{Dispatch} holds
        the registries of loaded bridges and implementations
        to be able to distinguish between them on the subsequent function calls.
        The Dispatch library communicates with a Bridge component
        via a defined interface.
  \item \emph{Bridge} provides the means to invoke a particular implementation
        as this is a language-specific component: for example,
        the Python bridge knows how to load Python modules, instantiate
        classes, and invoke functions.
  \item \emph{Interface} is an abstract component that specifies
        the interface (which is the same as for a Gateway,
        up to a hidden data structure that holds the state for non-object-
        oriented languages),
        and an \emph{Adapter} conform an existing implementation
        to this interface.
        Adapters do not have to be the part of the software package,
        although we bundle with the package several of them.
\end{enumerate}

\subsection{Supported library features}

\emph{MaRDI Open Interfaces} currently support programming languages
C, Julia, and Python on both user and implementation side.
As explained above, \emph{Converters} convert data to intermediate
C representation:
therefore, for passing data from C users no conversion is required, for Python the built-in module \texttt{ctypes}
is used, and Julia includes facilities to convert its data types
to corresponding C data types in its \emph{Base} package.

On the implementation side, \emph{Bridges} translate data to native data
types and invoke requested functions,
which in the case of the C Bridge requires use of \texttt{libffi}
for dynamic function calls,
while the Python Bridge and the Julia Bridge use corresponding
C APIs to instantiate embedded interpreters and pass data to invoked
functions.

When passing data between different languages,
we encode information about their types using
symbolic integer constants (their actual values are not relevant,
as long as they are consistent between different languages)
provided along with their description in~\autoref{tab:type-constants}.

\begin{table}
  \begin{center}
    \caption{Symbolic constants used in \emph{MaRDI Open Interfaces}
      to encode information about data types.\label{tab:type-constants}}
      \begin{tabular}{>{\raggedright\arraybackslash}p{0.30\textwidth}
          >{\raggedright\arraybackslash} p{0.45\textwidth}}
      \toprule
      Constant              & Description \\
      \midrule
      \texttt{OIF\_TYPE\_INT} &  32-bit integers \\
      \texttt{OIF\_TYPE\_F64} &  64-bit floating-point numbers \\
      \texttt{OIF\_TYPE\_ARRAY\_F64} &  arrays of 64-bit floating-point numbers \\
      \texttt{OIF\_TYPE\_STRING} &  strings \\
      \texttt{OIF\_TYPE\_CALLBACK} &  callback functions \\
      \texttt{OIF\_TYPE\_USER\_DATA} &  user-data objects of volatile type (\emph{opaque pointers} in C~parlance) \\
      \texttt{OIF\_TYPE\_CONFIG\_DICT} &  dictionary of key-value options pairs \\
      \bottomrule
    \end{tabular}
  \end{center}
\end{table}

For composite data types such as \texttt{OIF\_ARRAY\_F64}
and \texttt{OIF\_CONFIG\_DICT},
\emph{MaRDI Open Interfaces} include data structures in C,
so that interfaces in C and other languages mirror each other.
For example, for an interface with a function
requiring an array argument,
this array argument will be a NumPy array in Python, a built-in array
in Julia, and \texttt{OIFArrayF64} data structure in C.

\subsection{Data marshalling}

Function arguments during data marshalling are packed in three lists:
\begin{itemize}
    \item Immutable arguments that are read-only,
    \item Mutable arguments, which are used to write values to, however,
      the memory is provided by the caller.
      The only currently available mutable arguments are arrays
        (\texttt{OIF\_TYPE\_ARRAY\_F64})
        as the main data structure in scientific computing.
      This is necessary to avoid redundant memory allocations on the callee side,
      which can be especially expensive in case of the arrays.
    \item Output arguments, which are used to return data to the caller,
      in case when the memory is allocated by the callee.
      The use case for such arguments is, for example, when the callee needs
      to provide a string message to the caller.
      As the length of the string is not known in advance,
      it is the callee responsibility to allocate the memory
      and pass it to the caller.
\end{itemize}

Overall, data marshalling passes immutable and mutable arguments
by reference, hence, the memory is not copied,
which is important to minimize performance penalty associated
with crossing the inter-language boundary.
The memory is copied for the output arguments,
however, they are not used often, hence, their influence on performance
is negligible.

\subsection{Quality Assurance and Software Availability}

To ensure high reliability of \emph{MaRDI Open Interfaces},
we following common software-engineering practices such as
automated testing~\cite{Dubois2005, FritzEtAl2025}
and continuous integration~\cite{PetersEtAl2025}.

The test suite for the software package consists
of end-to-end tests written using \emph{Google Test},
\emph{Testing.jl},
and \emph{Pytest},
for different combinations of interfaces and implementations,
as well as a small number of unit tests.

Continuous integration occurs on \emph{GitHub Actions},
and serves as additional layer of quality assurance
by building the software package and running the test suite
from scratch on a separate machine.

Software package requires a UNIX-like operating system
and is available as source code on
\url{https://github.com/MaRDI4NFDI/open-interfaces}
under the BSD-2-Clause License.

%% file: 03-included-interfaces.tex
\section{Description of included interfaces}

\subsection{%
  Interfaces for solving initial-value problems
  for ordinary differential equations}

\emph{MaRDI Open Interfaces} include
an open interface \texttt{IVP}
for solving initial-value problems (IVP)
for ordinary differential equations (ODEs),
namely, for problems of the form:
\[
  y'(t) = f(t, y), \quad y(t_0) = y_0,
\]
where $y(t) \colon \R \to \R^{n}$, $f(t, y) \colon \R \times \R^{n} \to \R^{n}$,
$y_{0}$ is the initial system state at time $t_{0}$.

This interface as well as supported implementations
is described in details with example usage
in our previous work~\cite{KabanovEtAl2025}.

\subsection{Interface for solving nonlinear optimization problems}

Another interface, \texttt{optim}, included in \emph{Open Interfaces} is
for solving nonlinear optimization problems,
and it currently supports unconstrained problems of the form
\begin{align}
  \argmin_{x \in \R^n } \quad & f(x; \alpha)
\end{align}
where the objective function  \( f: \R^{n} \times \mathbb A \to \R \)
is additionally parameterized with some context data \(\alpha \in \mathbb A\).

The interface for such problems in general requires the following operations.
One needs to be able to set an initial guess for the sought-for parameter vector \(x\), 
as the algorithms for nonlinear optimization are iterative.
Besides, as normally the objective function requires additional data,
we need to be able to pass them to the objective function.
Additionally, depending on the algorithm used, we might need to provide
not only the objective function callback but also the gradient of the objective function.
Even more, we need to be able to set the optimization algorithm and its parameters:
in general, different algorithms and implementations use different stopping criteria and other parameters,
as will be seen further in the Section~\ref{sec:examples}.
Finally, different algorithms and implementations use different status codes and result messages
to report the result of the optimization process, so it is useful to pass them back to the user
as is.

Taking into the account the above considerations, the interface
for unconstrained nonlinear optimization has currently
the following function calls available
(again, in a pseudo-C syntax):
\begin{minted}[escapeinside=||]{C}
// Set an initial guess for the sought-for parameter vector |\(x\)|.
OIF_TYPE_INT set_initial_guess(OIF_TYPE_ARRAY_F64 x0);
// Set the user-provided context |\( \alpha \)| for |\(f\)|.
OIF_TYPE_INT set_user_data(OIF_TYPE_USER_DATA user_data);
// Set the callback for |\( f \)|.
OIF_TYPE_INT set_objective_fn(OIF_TYPE_CALLBACK objective_fn);
// Set the callback for |\( \nabla_x f \)|.
OIF_TYPE_INT set_grad_fn(OIF_TYPE_CALLBACK grad_fn);
// Set the optimization algorithm and its parameters.
OIF_TYPE_INT set_method(
    OIF_TYPE_STRING method_name,
    OIF_TYPE_CONFIG_DICT method_params
);
// Minimize |\( f (x) \)| and write the resultant |\( x \)| into `out_x`.
(OIF_TYPE_INT, OIF_TYPE_STRING) minimize(
    OIF_TYPE_ARRAY_F64 out_x
);
\end{minted}

In the current version of \emph{Open Interfaces},
the following optimization-interface adapters are available:
\begin{itemize}
  \item The \texttt{scipy\_optimize} adapter
        for the \texttt{optimize} subpackage of
        \emph{SciPy}~\cite{VirtanenEtAl2020},
  \item The \texttt{optim\_jl} adapter
        for the \emph{Optim.jl} package~\cite{MogensenRiseth2018}.
\end{itemize}

%% file: 04-examples.tex
\section{Example usage}\label{sec:examples}

We demonstrate the usage of \emph{MaRDI Open Interfaces}
with a~standard problem in nonlinear optimization~--- finding a minimum of the Rosenbrock function:
\begin{align}
  \label{eq:min-rosen}
    \argmin_{x\in\R^n} \quad &f(x; a), \\
    \label{eq:min-rosen-fn}
    \mathrm{with}  \quad &
    f(x; a) = \sum_{i=1}^{n-1} a {\left( x_{i+1} - x_{i-1}^2 \right)}^2
      + {\left(1 - x_i\right)}^2,
\end{align}
where for this example we set $a = 10$ and $n = 5$.

It is known that for \(n = 5\), this function has the global minimum \(f(x_\mathrm{opt}) = 0\),
where \(x_\mathrm{opt} = (1, 1, 1, 1, 1)^\mathrm{T}\).

The gradient of this function is defined by
\begin{equation*}
\begin{aligned}
\frac{\partial f}{x_1} &= -4a x_1 \left(x_2 - x_1^2\right) - 2(1-x_1), \\
\frac{\partial f}{x_i} &= 2a \left(x_i - x_{i-1}^2\right) \\
                       &\phantom{=}  -4a x_i \left(x_{i+1} - x_{i}^2\right) - 2(1-x_i), \quad i = 2, \dots, n-1,\\
\frac{\partial f}{x_n} &= 2a \left(x_n - x_{n-1}^2\right).
\end{aligned}
\end{equation*}

We use implementations of the gradient-free Nelder-Mead simplex method~\cite{NelderMead1965, GaoHan2012}
and the gradient-based BFGS method~\cite{Broyden1970a, Broyden1970b, Fletcher1970, Goldfarb1970, Shanno1970}
from \emph{SciPy} and \emph{Optim.jl} packages
to solve this optimization problem.
For the Nelder-Mead method, as it is a simplex method,
the stopping criterion for the optimization process is
based on the function values on the vertices of the simplex being close to each other
within tolerance \texttt{funtol}
and differs for two implementations:
in \emph{SciPy} the condition is based on the ``\(\infty\)-norm''
of the function values on the vertices of the simplex~\cite{GaoHan2012}:
\[
\max_{2\leq i \leq n+1} \left|f_i - f_1\right| \leq \texttt{funtol}
\]
while in the implementation in \emph{Optim.jl},
the stopping criterion is taken from the original paper~\cite{NelderMead1965},
and based on the ``standard error'' of the function values on the vertices
of the simplex:
\[
\sqrt{\frac{\sum_{i=2}^{n+1} |f_i - f_1|}{n}} \leq \texttt{funtol}.
\]
In our example here, we set \(\texttt{funtol} = 10^{-11}\).

For the gradient-based BFGS method, both implementations
allow the same stopping criterion: \(\infty\)-norm of the gradient vector
below a user-defined tolerance value \texttt{gtol},
which we set to \(10^{-8}\).

Additionally, we set the \texttt{linesearch} parameter for the BFGS implementation
from \emph{Optim.jl} to \texttt{StrongWolfe}~\cite[Ch.~3]{NocedalWright2006}
to match the hardcoded line-search algorithm in \emph{SciPy} for the BFGS method.
In that sense, gradient-based algorithms in \emph{Optim.jl} provide
more flexibility, as a~line-search strategy is passed explicitly to the solver.

As an initial guess for the optimization process, we set
\begin{equation}
    x_0 = {(3.14, 2.72, 6.18, 9.81, 8.31)}^\mathrm{T}.
  \label{eq:min-x0}
\end{equation}

A shortened version of the Julia script used for these simulations
is provided in the following listing.

\begin{minted}[linenos]{julia}
using OpenInterfaces.Interfaces.Optim

function rosenbrock_objective_fn(x, a)
    return sum(
      a * (x[2:end] - x[1:(end-1)] .^ 2.0) .^ 2.0 +
      (1 .- x[1:(end-1)]) .^ 2.0
    )
end

function rosenbrock_grad_fn(x, grad_f, a)
    xi = @view x[1:(end-1)]
    xip1 = @view x[2:end]

    grad_f[1:(end-1)] .= -4.0 .* a .* xi .* (xip1 .- xi .^ 2.0)
    grad_f[1:(end-1)] .-= 2.0 .* (1.0 .- xi)
    grad_f[end] = 0.0
    grad_f[2:end] .+= 2.0 .* a .* (xip1 .- xi .^ 2.0)
    return 0
end

x0 = [3.14, 2.72, 6.18, 9.81, 8.31]
user_data = 10

impl = "optim_jl"  # or "scipy_optimize"
method = "BFGS"    # or "NelderMead"
linesearch = "StrongWolfe"

s = Optim.Self(impl)
Optim.set_initial_guess(s, x0)
Optim.set_user_data(s, user_data)
if impl == "scipy_optimize"
    if method == "NelderMead"
        Optim.set_method(
          s, "nelder-mead", Dict("fatol" => 1e-11)
        )
    elseif method == "BFGS"
        Optim.set_method(s, "BFGS", Dict("gtol" => 1e-8))
    end
elseif impl == "optim_jl"
    if method == "NelderMead"
        Optim.set_method(
          s, "NelderMead", Dict("g_abstol" => 1e-11)
        )
    elseif method == "BFGS"
        Optim.set_method(
          s, "BFGS",
          Dict("g_abstol" => 1e-8, "linesearch" => linesearch)
        )
    end
end
Optim.set_objective_fn(s, rosenbrock_objective_fn)
Optim.set_grad_fn(s, rosenbrock_grad_fn)

status, message = Optim.minimize(s)
@show status
@show message
@show s.x
\end{minted}
The full version can be found in the project repository
in the file\\
\texttt{examples/lang\_julia/call\_optim\_rosenbrock.jl}.

Results of solving the optimization
problem~\eqref{eq:min-rosen}--\eqref{eq:min-rosen-fn}
using these algorithms
are provided in~\autoref{tab:results}.
The results demonstrate the following.
First, it is clear that the gradient-based BFGS method converges faster,
with significantly fewer iterations than the derivative-free Nelder--Mead method,
although, of course, the former requires the gradient function and its evaluation,
which can be costly in high dimensions.
Not only that, but the obtained final objective value \(f(x_*)\) is much closer
to the theoretical optimal value (of zero for the function~\eqref{eq:min-rosen})
for the BFGS method than for the Nelder--Mead method.
Second, we can see that although the same methods are used
for comparison, there is a significant variance in the number of required
iterations, function and gradient evaluations between different implementations
of the same numerical algorithms.

\begin{table}
\caption{
  Results of running the Nelder--Mead and~BFGS optimization algorithms
  as implemented in the packages \emph{SciPy} and \emph{Optim.jl}
  via \emph{MaRDI Open Interfaces} on the problem~\eqref{eq:min-rosen}-\eqref{eq:min-rosen-fn}
  with the initial condition~\eqref{eq:min-x0}
  and the true solution~\(x_\mathrm{opt}\).\label{tab:results}
}
\begin{center}
\begin{tabular}{l c c c c}
\toprule
          & \multicolumn{2}{c}{Nelder--Mead} & \multicolumn{2}{c}{BFGS} \\
            \cmidrule(lr){2-3} \cmidrule(lr){4-5}
Statistic &  \emph{SciPy} & \emph{Optim.jl} & \emph{SciPy} & \emph{Optim.jl} \\
\midrule
Final \(f(x_*)\)              & \num{9.62e-13} & \num{1.76e-12} & \num{5.15e-19} & \num{6.81e-19} \\
\(\norm{x_* - x_\mathrm{opt}}_\infty < 10^{-5}\) & yes            & yes            & yes            & yes            \\
No.\ of iterations                             & 515            & 446            & 69             & 46             \\
No.\ of evaluations of \(f\)                   & 839            & 760            & 81             & 131            \\
No.\ of evaluations of \(\nabla f\)            & -              & -              & 81             & 99             \\
Final \(\norm{\nabla f(x_*)}_\infty\) & -              & -              & \num{5.36e-9}  & \num{3.65e-9}        \\
\bottomrule
\end{tabular}
\end{center}
\end{table}

As one can see from these example, \emph{MaRDI Open Interfaces}
provide computational scientists the ability to try different solvers
for the same problem type without writing cross-language bindings or adapting their code
to different solver interfaces,
while at the same time keeping the project-specific code in their language of choice.
In this case, we used Julia as the primary language for the experiment script,
which calls implementations of the optimization algorithms
from the Python and Julia packages without rewriting the code.
However, as one can see even in this simple example,
deeper knowledge on the implementation details
(such as a stopping criterion for an iterative algorithm)
is still helpful, especially in benchmarking tasks,
to compare different implementations on equal grounds.